\documentstyle[pre,preprint,aps,amssymb,rotating,epsfig]{revtex}

\tightenlines

\newcommand{\be}{\begin{equation}}
\newcommand{\en}{\end{equation}}
\newcommand{\bea}{ \begin{eqnarray}}
\newcommand{\ena}{\end{eqnarray} }
\renewcommand{\b}{\beta}
\newcommand{\e}{\eta}
\newcommand{\s}{\sigma}
\newcommand{\p}{\psi}
\renewcommand{\d}{\delta}
\newcommand{\ra}{\rightarrow}

\begin{document}

\title{First order phase transition in a nonequilibrium growth process}
\author{Lorenzo Giada$^{*\dagger}$, Matteo Marsili$^*$}
\address{$^*$INFM, Trieste-SISSA Unit, Via Beirut 2-4, Trieste I-34014 \\ 
 $^\dagger$International School for Advanced Studies (SISSA/ISAS), 
Via Beirut 2-4, Trieste I-34014 }
\date{\today}
\maketitle

\begin{abstract}
We introduce a simple continuous model for nonequilibrium surface growth. 
The dynamics of the system is defined by the KPZ equation with a Morse-like 
potential representing a short range interaction between the surface and the 
substrate. 
The mean field solution displays a non trivial phase diagram with a first 
order transition between a growing and a bound surface, associated with a 
region of coexisting phases, and a tricritical point where the transition 
becomes second order. 
Numerical simulations in 3 dimensions show quantitative agreement with mean
 field results, and the features of the phase space are preserved even in 
2 dimensions.
\end{abstract}

\section{Introduction}

Recently Hinrichsen \emph{et.~al.~}\cite{hlmp2} have introduced a discrete model for nonequilibrium surface growth that displays a first order unbinding transition in a region of the parameter space where there is phase coexistence between a growing and a bound surface.
They used a restricted solid on solid dynamics, and distinguished the probability of adsorption of particles at the substrate from that at higher layers. This in order to take into account short range interactions between the substrate and the surface.

In the same spirit we 
study a continuum growth model for an interface interacting with a substrate through a short range attractive potential. The interface separates a solid from a vapor phase, and is driven by a difference in chemical potential due, for example, to a flux of incoming particles, as in Molecular Beam Epitaxy. As the driving force increases the interface eventually detaches from the substrate and grows indefinitely according to KPZ \cite{KPZ} dynamics. We will focus on the nature of the transition from a bound to a moving phase. We resort to numerical simulations and to a mean field approach first introduced in \cite{marsili} for the KPZ equation. 

A similar unbinding process of a surface from a wall has been studied by Mu\~noz and Hwa \cite{DNA}. They considered a KPZ \cite{KPZ} equation to which they added strictly attractive or repulsive interactions, and identified a second order unbinding transition for which they calculated some critical exponents. 

We will show how the short range attractive potential changes the nature of the phase transition, which becomes first order for an entire range of values of the parameters of the model. The second order transition is recovered only for larger values of the noise, or in the case of a very short ranged potential, but the critical exponent is still different from that of the simple repulsive wall case.

Our model is an example of continuous nonequilibrium system with noise. In the context of noise induced phase transition \cite{horst} this kind of systems has been studied by Van den Broeck \emph{et al.~}\cite{toral,Kawai}, who showed that multiplicative noise is essential for the transition to take place, since it can trigger instabilities in the short time dynamics of the system. 
On the other hand systems with multiplicative noise are characterized by a transition to an absorbing phase \cite{grim}, a state from which the system cannot escape, characterized by a null value of the order parameter. The phase diagram of this class of systems has been extensively studied in recent years \cite{MN,MN2,mun}. 
Most of these works were concerned with second order transitions, but an example of first order has been studied by M\"uller \emph{et al.~}\cite{first} with the help of mean field techniques and numerical simulations.

\section{The Model}
With the term \emph{surface} we identify the interface between the solid and vapor phases of some substance. 
In what follows we want to give a statistical physics description of the interactions that may take place between the vapor particles as they are depositing on some kind of substrate.

First of all a note on terminology: an interaction is called attractive \emph{for the surface} when there is some force that repels incoming particles, so that the surface is not growing; On the contrary, attractive interaction between the substrate and the particles will let the surface grow and eventually detach from it.

In our model the system is described by a height field $h(x)$ defined on a $d$-dimensional continuous substrate, which evolves according to a Langevin equation of the KPZ type. The latter is chosen in order to have a system out of equilibrium.

The Langevin equation for the field $h(x)$ is obtained in the usual way:
\be
\partial_t h(t) = D\left[\nabla^2 h - (\nabla h)^2 \right]+r- \frac{\d V[h]}
{\d h}+ \s\e \quad ,
\en
where $D$ is some diffusion coefficient, $r$ is the driving force due to the incoming particles, and the nonlinear term $(\nabla h)^2$ characterizes the nonequilibrium KPZ-like \cite{KPZ} dynamics, and allows for the growth along the direction normal to tilted areas of the surface.
This term breaks the up-down symmetry of the system and therefore defines a preferential direction for growth (see \emph{e.~g.~}\cite{bar}).

The field $\e(x,t)$ is a white Gaussian noise with the following properties:
\bea\label{eq:nois}
&&\langle \e(x,t)\rangle = 0 \nonumber \\
&&\langle  \e(x,t) \e(x',t') \rangle = \d(x-x') \d (t-t') \quad .
\ena

 We model the interaction of the substrate with the surface introducing a potential $V[h]$ which has the shape of the Morse potential (see Fig.~\ref{fig:morse}):
\be\label{eq:pot}
V_M[h]=A (1-e^{- \b h}e^{\b h_0})^2 \quad ,
\en
where $A$ is the depth of the well, $\b$ the hardness of the repulsive wall that mimics the substrate, and $h_0$ the position of the local minimum, which can be set to $0$ without loss of generality.

Note that the driving force $r$ can be included in the potential as a linear term in the height field: $-rh(x)$. 
The parameter $r$ is the essential one: if it is negative the shape of $V$ is a well with an exponential wall on the small-$h$ side, whereas if it is positive the minimum close to the repelling wall is only a local one (Fig.~\ref{fig:morse}). Particles incoming from above in this case have to trespass a potential barrier (\emph{i.~e.~}the local maximum of $V$) 
in order to arrive close to the substrate. This is equivalent to a repulsion for particles, 
or an attraction for the surface. Physically this term has a precise meaning, since it is 
due to the difference in chemical potential between the solid and vapor phases, and determines a drive of the interface toward higher or lower values, depending on its sign.

In the dynamics of growing surfaces this term is usually eliminated via the transformation $h \ra h-rt$, which correspond to viewing the system from a moving reference frame. As one can see the potential in Eq.~(\ref{eq:pot}) is not invariant for such a transformation, 
nor for the more general $h \ra h+\d h$. Notice that the KPZ equation is invariant for both these transformations.
In the present case the consequence is that we can uniquely define the mean position of the surface with respect to the substrate, and its state of motion. It is therefore meaningful to study the unbinding of the surface from the substrate.

By inserting the Morse potential one obtains:
\be\label{eq:lang}
\partial_t h(t) = r+2\b A(e^{-2\b h}-e^{-\b h})+D\left[\nabla^2h-(\nabla h)^2\right] +\s\e \quad .
\en

By performing a Hopf-Cole transformation $\p=e^{-h}$ we map Eq.~(\ref{eq:lang}) into a Langevin equation with multiplicative noise; this suggests us the order parameter to be used in the search for a phase transition, namely the spatial average of $\p$, $m\equiv\langle\p\rangle$:
\be\label{eq:psi}
\partial_t \p(t)= (-r+\s^2 /2)\p - 2\b \left[\p^{2\b+1}- \p^{\b +1}\right]+ \nabla^2 \p +  \s\p\e \quad .
\en
Notice that we have used the Ito calculus in doing the transformation, and
we have been able to set $A$ and $D$ equal to one via a suitable rescaling of space and time.

Qualitatively we can have two possible situations in the long time limit of the stationary solution of the equation. Either $m\ra0$, which corresponds to a diverging value of $\langle h\rangle$, \emph{i.~e.~}to a moving (growing) surface; or $m\not=0$, corresponding to the surface being bound to the substrate at some average height. The transition between the two phases is controlled by the parameter $r$. 
Namely, as one can see from Eq.~(\ref{eq:lang}), for large negative values of $r$ the surface is pushed down against the substrate, whereas for large positive values the surface is pulled away from it. It is the balance between the force $r$ and the repulsion from the exponential wall $e^{-2\b h}$ that determines the equilibrium position of the surface. Notice finally that in the language of multiplicative noise systems the state with $m=0$ is the absorbing one, since the surface can never invert its average velocity and go back toward the substrate.


An equation similar to (\ref{eq:psi}) has been studied by Mu\~noz and Hwa in \cite{DNA} in the context of multiplicative noise processes. The system had a soft lower repulsive wall represented by a term $-p \p^{p+1}$ in the Langevin equation, the hard wall limit corresponding to $p \ra \infty$:
\be\label{eq:DNA}
\partial_t \p= \nabla^2\p +r\p -p\p^{p+1}+\p\s\e  \quad .
\en
 The authors found a second order phase transition between a bound and a moving phase. From scaling arguments, the exponent that controls the vanishing of the order parameter close to the transition is found to be $\b_r=(2-z)/(2z-2)$, where $z$ is the KPZ dynamical exponent. Numerical simulations confirmed this result.

\section{Mean Field Approximation }

To proceed further with the analytical solution of the model we notice that dimensional analysis performed on Eq.~\ref{eq:psi} along the lines of \cite{MN} shows that the upper critical dimension for the depinning transition is 2.
This means that mean field results should be correct in 3 dimensions, and provide a reasonable estimate (up to logarithmic corrections) in two dimensions, \emph{i.~e.~}for real surfaces.

First of all we perform a discretization in space, by inserting a (hyper)cubic lattice, and define the variables $\p( \mathbf{x}$$ ,t) \ra \p(x_i,t)\equiv\p_i(t) $. The discretized Laplacian is: 
\be\label{eq:lap}
\nabla^2 \p_i = \frac{1}{d}\sum_{j\in nn(i)}(\p_j-\p_i)\quad,
\en
where the sum runs over the nearest neighbor of the site $i$.
The mean field approximation we are going to use consists \cite{marsili,mun} in the following expression for the discretized Laplacian:
\be\label{eq:mean}
\frac{1}{d}\sum_{j\in nn(i)}(\p_j-\p_i)\simeq \langle\p\rangle - \p_i \quad.
\en
This is equivalent to say that the number of dimensions is very large, or that the number of neighbors (coordination of the lattice) is infinite, or again that the interaction is infinite range.

Equation (\ref{eq:psi}) then becomes:
\be\label{eq:mfapsi}
\partial_t \p_i(t) = (-r+\s^2 /2 -1)\p_i - 2\b \left[\p_i^{2\b+1}- \p_i^{\b +1}\right]+
\langle \p \rangle - \s\p_i\e_i \quad , 
\en
from which, going back to the field $h$, the discretized version of Eq.~(\ref{eq:lang}) becomes:
\be\label{eq:mfa}
\partial_t h_i(t) = (r+1) + 2\b e^{-\b h_i}(e^{-\b h_i}-1) - \langle e^{-h} \rangle e^{h_i} + \s \e_i \quad .
\en

\subsection*{The Phase Transition} 
\noindent
We obtain the probability distribution of the stationary solution of Eq. (\ref{eq:mfa}) by solving the associated Fokker-Planck equation. The stationary solution is:
\[
P_{st}(h) = {\mathcal N} e^{-\frac{2V_{eff}(h)}{\s^2}}\quad ,
\]
where $\mathcal N$ is the normalization factor, and the effective potential is in the present case:
\be
V_{eff}(h) = -(r+1)h+e^{-2\b h}-2e^{-\b h}+\langle e^{-h}\rangle e^h \quad .
\en
This expression for the probability distribution depends on the value of the order parameter, which must be determined self-consistently. We have to solve therefore the following mean-field equation:

\bea
\label{eq:self}
m = \langle e^{-h}\rangle &=& \int e^{-h} P_{st}(h)dh
\equiv F(m)=\frac{g_J(m)}{g_{J-1}(m)}\\
g_j(m)&\equiv &\int_{-\infty}^\infty dh e^{-(j+1)h}e^{-\frac{2}{\s^2}
e^{-2\b h}}e^{\frac{4}{\s^2}e^{-\b h}}e^{-\frac{2m}{\s^2}e^h} \quad , 
\ena
where $J=-2(r+1)/\s^2$.

Plotting $m$ and $F(m)$ on the same graph, it is easy to see that Eq.~(\ref{eq:self}) can have one, two or three solutions in $m$, according to the value of $r$. The possible situations are represented in Fig.~\ref{fig:self}. The stability of the solutions can be investigated making use of the results of Shiino \cite{shi}. Namely, the solution can be stable only when $F(m)$ crosses the line $m$ from above.
In our case we have that for $r$ larger than some $r_{c_2}$
the only solution (therefore a stable one) is $m=0$; for $r$ below some value which will be determined in a moment, there are only 2 solutions, the stable one being $m\not=0$; finally for intermediate values of $r$ there are three solutions, two of which look stable. Notice that $m=0$ remains always a solution;
it becomes stable only when $r$ reaches the value for which $F'(m=0)=1$. This value is found analytically to be: $r_c=\frac{\s^2}{2}$. This number is also the lowest value of $r$ for which three solutions exist. 

The situation just described represents a typical case of first order phase transition, and is the same as found in \cite{first}. The value of the order parameter $m$ remains different from zero as long as this is the only stable solution. Then for $r>r_c$ there is a coexistence region, where two phases are stable, and the final state of the system is determined by the initial conditions. 
Finally there is the value  $r=r_{c_2}$ above which the order parameter can be only zero. If we remind the definition of $m$ we understand that this phase transition corresponds to the unbinding of the surface from the substrate, $\langle e^{-h}\rangle \ra 0$ meaning $\langle h \rangle\ra \infty$. We remark that with the term \emph{coexistence} we do not mean that both phases are present at the same time, rather that depending on the initial conditions both are possible stationary state of the system.
In Fig.~\ref{fig:ist} is shown an example of this phenomenon. In the coexistence region an interface which is initially close to the substrate will remain bounded to it whereas if the interface is initially far it will move further away from the substrate with a velocity $v>0$.

\subsection*{The Moving Phase} 
\noindent
It is also possible to determine the velocity of the surface when it is away from the substrate. To do this we define a field $\phi_i(t)$ for the fluctuations of $h$ around its mean position: $\phi_i(t)=h_i(t)-vt$. Here $v$ is the mean velocity of the surface obtained from Eq.~(\ref{eq:mfa}): 
\[
v=\partial_t\langle h_i\rangle=r+1-\langle e^{-h}\rangle\langle e^{h}\rangle\quad,
\]
where we could neglect the exponential terms in the potential, since they are vanishingly small in the moving phase. 
The field $\phi_i(t)$ obeys in this phase to the following Langevin equation:
\[
\partial_t \phi_i=\partial_t h_i -v=(r+1-v)-\langle e^{-h}\rangle e^{h_i}+\s\e_i\quad .
\]
The stationary probability distribution from the associated Fokker-Planck equation is as usual (see \emph{e.~g.~}\cite{gard}):
\be\label{eq:vel}
P_{st}(\phi)={\mathcal N} e^{-\frac{2V_{eff}(\phi)}{\s^2}}={\mathcal N}e^{\alpha\phi -\gamma e^\phi} \quad,
\en
with $\alpha=\frac{2}{\s^2}(r+1-v)$ and $\gamma=\frac{2}{\s^2}\langle e^{-\phi}\rangle $. 
The self-consistency relation for $\langle e^{-h}\rangle$ is:
\[
\langle e^{-h}\rangle=\gamma \frac{1}{\alpha-1}\quad ,
\]
from which we have that $v=r-\frac{\s^2}{2}$. The condition $v>0$ shows that $r_c=\s^2/2$, as anticipated.
In the next section we will see that for large values of the noise strength $\s$ and of $\b$, the unbinding transition becomes second order. In that part of the phase diagram we can define in the vicinity of the transition a velocity critical exponent $\theta$ \cite{bar}: $v\sim (r-r_c)^\theta$. From the previous discussion we see that $\theta=1$.

\section{Numerical Findings}

To study further the phase diagram of the model in mean field approximation we solve numerically Eq.~(\ref{eq:self}). In particular we have determined the value $r_c^*$ for which its two non-zero solutions coalesce with the null one. Analytically this point corresponds to a negative second derivative of $F(m)$ at the critical point $r_c$.
 In Figs.~\ref{fig:ph2} and \ref{fig:ph1} are shown two typical cuts of the phase diagram at fixed values of the control parameters $\s$ and $\b$ respectively. One can see that as the value of one of the parameters increases while the other remains fixed, the coexistence region becomes smaller, and eventually disappears for some $\b^*(\s)$ or $\s^*(\b)$. At the same time the unbinding transition becomes second order, thus defining a line of tricritical points in the phase diagram ($\s$,$\b$) of the system.
A numerical estimate of the exponent $\b_r$ governing the transition (see \emph{e.~g.~}\cite{mun}) reveals that, at least for large values of $\b$, we have:
\be
m \sim |r-r_c|^{\b_r}  \sim |r-r_c|^{-\s^2/2}
\en

This can be checked analytically by taking the limit for $\b\ra\infty$ at the critical point in Eq.~(\ref{eq:self}). The condition to have a second order transition is that $F''(m)<0$ at $r_c$; in this case there can be no coexistence region, since $r_{c_2}=r_c$.
With the same notation of Eq.~(\ref{eq:self}) we have:
\begin{eqnarray}
F^{'}(m)&=&-\frac{2}{\s^2}\left(1-\frac{g_Jg_{J-2}}{g_{J-1}^2}\right)\\
F^{''}(m)&=&\left(\frac{2}{\s^2}\right)^2\left[-\frac{g_{J-2}}{g_{J-1}}-\frac{g_Jg_{J-3}}{g_{J-1}^2}
+2\frac{g_Jg_{J-2}^2}{g_{J-1}^3}\right] \quad .
\end{eqnarray}
Then we make the change of variable $x=2m/\s^2 e^h$, to get:
\begin{eqnarray*}
g_j(m)&=&\left(\frac{2m}{\s^2}\right)^{j+1}\int_{\frac{2m}{\s^2}}^\infty dx x^{-(j+2)}e^{-x} \\
&=&\left(\frac{2m}{\s^2}\right)^{j+1}\left[\Gamma(-j-1)-\int_0^{\frac{2m}{\s^2}}dx x^{-(j+2)}e^{-x} 
\right] \quad .
\end{eqnarray*}
Since we are looking for a second order transition, we need to estimate the small $m$ limit of $g_j(m)$.
Therefore we can expand $e^{-x}$ in the integral, and solve it explicitly to any order in $m$.
At the critical point $r_c=\s^2/2$ we find that:
\[
F^{''}(m)=-\frac{\frac{2}{\s^2}+1}{\Gamma(\frac{2}{\s^2}+1)}\left(\frac{2m}{\s^2}\right)
^{\frac{2}{\s^2}-1} < 0\quad ,
\]
which means that the transition is always second order at large $\b$, and that the $m$-expansion of $F(m)$ close to the critical point can be written as:
\[
F(m)=(1+\delta r)m+Cm^{\frac{2}{\s^2}+1} \quad ,
\]
where $\delta r=r_c-r$, and $C$ is some constant. Solving the mean field equation $F(m)-m=0$ gives:
\be
m \sim \left(\delta r\right)^\frac{\s^2}{2} \quad ,
\en
in agreement with the numerical value.

We note that the latter calculation can be repeated for the model of \cite{DNA} Eq.~(\ref{eq:DNA}), to give the same result, thus implying that in the mean field limit of a system with a hard wall the phase transition is governed by a continuous exponent. To our knowledge this finding was not previously reported. A different value for the $\b_r$ exponent, namely $\b_r=1/p$, is given in \cite{mun}, where the mean field approximation Eq.~(\ref{eq:mean}) is also compared with different results from field theoretical calculations ($\b_r=1$). However it is not clear whether the hard wall limit can be obtained from the aforementioned calculations.

Another feature of the phase diagram is that in the limit of small $\s$ $r_{c2} \ra \b/2$
, but we were unable to derive this result analytically from Eq.~\ref{eq:self}.

We check if there is any functional relation between $\b^*$ and $\s^*$. Figure \ref{fig:tric} shows a log-log plot of the tricritical points together with the linear fitting. The value of the slope is in this case about $-4$, which means that $\b^* \sim \s^{*-4}$, or that $\b^* \sim r_c^{*-2}$. Notice however that this linear fitting is a fairly poor approximation, and there is no reason why we should expect a priori any simple functional relation between these two quantities.

We have also checked with simulations on a finite number of sites that these results are the same as those obtained from Eq.~(\ref{eq:mfa}).
With the help of these simulations we could see that hysteresis associated with the first order transition and the coexistence region does indeed occur.

\section{Results in Finite Dimension}

To perform numerical simulation of Eq.~(\ref{eq:lang}) in finite dimensions we introduced as usual a (hyper)cubic lattice on the space of coordinates, and used for the discretized Laplacian the expression in Eq.~(\ref{eq:lap}). Going back to the $h$ field we obtained the following Langevin equation on a lattice:
\be\label{eq:disc}
\partial_t h_i=r+1+2\b e^{-\b h_i}(e^{-\b h_i}-1) -e^{h_i}\frac{1}{d}
\sum_{j\in nn(i)} e^{-h_j} + \s \e_i \quad .
\en
To have a better accuracy we chose to apply the Heun method for solving stochastic differential equations (see \emph{e.~g.~}\cite{heun,heun2}). We run the simulations long enough for the system to reach its stationary state, and then measured the order parameter at intervals larger than some estimated autocorrelation time.

The results from three-dimensional simulations are in quantitative agreement with the mean field analysis, thus confirming that the upper critical dimension for this class of systems is $2$.

It is now interesting to see how this analysis is modified in the two-dimensional case, the physical one, when in the renormalization group sense only logarithmic corrections are expected. 
We simulated systems on a square lattice with linear size $L$ up to 60. A cut of the phase space is shown in Fig.~\ref{fig:2d}. It is easily seen how the main features of the phase diagram remain unaltered in 2 dimensions; namely we can still identify a first order phase transition with a coexistence region where the system displays hysteresis (Fig.~\ref{fig:ist}), and a crossover to a second order transition for large values of the noise.

Similar results are obtained even in 1 dimension. It must be noticed how the coexistence region becomes smaller as one goes to low dimensionality, but is still present. 

\section{Discussion and Conclusions}

In a sense the combined effect of the attracting potential and of the noise is similar to that of a superposition of quenched an thermal noise (\emph{cfr.} \cite{bar}). Namely in both situations we have a force opposing the growth that depends on the height of the surface, and a noise term that randomly modulates in time and space such interaction. This is why we can use the quenched noise terminology and define \emph{e.~g.~}the velocity exponent. Of course it is the attractive nature of the Morse potential that gives the transition its first order character, and we expect similar results for other short range attractive potential as well.

We have shown how a simple modification of the nonlinear part of the KPZ equation can describe the dynamics of a surface that interacts with the substrate via a short range interaction. The mean field analysis has revealed that a first order unbinding transition takes place for a large range of the control parameters, and that in a whole region of the phase diagram the system's stationary position depends on the initial conditions (hysteresis). Furthermore a  crossover to a second order transition takes place for large values of the control parameters, and the exponent controlling the vanishing of the order parameter depends continuously on the strength of the noise.

The same situation is recovered in simulations above and below the upper critical dimension $d_c=2$

\subsubsection*{Acknowledgments}
\noindent
We acknowledge stimulating discussions with M.~A.~Mu\~noz, H.~Hinrichsen, D.~Mukamel, A.~De Martino.

\begin{figure}
\caption{Morse potential including the driving force $-rh$.}
\label{fig:morse}
\end{figure}
\begin{figure}
\caption{Graphical solution of the self consistence equation for $\b=1$, $\s=0.4$.}
\label{fig:self}
\end{figure}
\begin{figure}
\caption{Hysteresis curve obtained from mean field (MF) and 2-$d$ simulations.
Here \emph{close} and \emph{distant} indicate the initial position of the surface with respect to the substrate.}
\label{fig:ist}
\end{figure}
\begin{figure}
\caption{The mean field phase diagram for fixed $\s=1$. The continuous line corresponding to $r_c=\s^2/2$ merges with the dotted line $r_{c_2}$ at the tricritical point ($\times$). For larger $\b$ the transition is second order (dashed line).}
\label{fig:ph2}
\end{figure}
\begin{figure}
\caption{The mean field phase diagram for $\b=1$. The continuous line corresponding to $r_c=\s^2/2$ merges with the dotted line $r_{c_2}$ at the tricritical point (+). For larger $\s$ the transition is second order (dashed line).}
\label{fig:ph1}
\end{figure}
\begin{figure}
\caption{The relation between $\b^*$ and $\s^*$ in a log-log plot}
\label{fig:tric}
\end{figure}
\begin{figure}
\caption{A cut of the phase diagram for the 2-dimensional system at fixed $\b=1$. \emph{Close} and \emph{distant} indicate the initial position of the surface with respect to the substrate.}
\label{fig:2d}
\end{figure}


\begin{thebibliography}{910}
\bibitem{hlmp2}H. Hinrichsen, R. Livi, D. Mukamel, and A. Politi,
Phys. Rev. E {\bf 61}, R1032 (2000)

\bibitem{KPZ}
M. Kardar, G. Parisi and Y. C. Zhang,
Phys. Rev. Lett. {\bf 56}, 889 (1986).

\bibitem{marsili}
M.~Marsili and A.~J.~Bray,
Phys. Rev. Lett. {\bf 76}, 2750 (1996)

\bibitem{DNA} M. A. Mu{\~{n}}oz and T. Hwa,
 Europhys. Lett. {\bf 41}, 147 (1998).

\bibitem{horst}W. Horsthemke e R. Lefever, {\it Noise Induced Transitions},
Springer Verlag, Berlin, 1984

\bibitem{toral} C. Van den Broeck, J. M. R. Parrondo and R. Toral,
Phys. Rev. Lett. {\bf 73} 3395 (1994)

\bibitem{Kawai}  C. Van den Broeck, J. M. R. Parrondo, R. Toral
and R. Kawai, Phys. Rev. E { \bf 55}, 4084 (1997).

\bibitem{grim}G. Grinstein, M. A. Mun\~oz,
{\it The Statistical Mechanics of Absorbing States}, in {\it Fourth
Granada Lectures in Computational Physics}, Ed. P. Garrido and
J. Marro, Springer (Berlin),
Lecture Notes in Physics, {\bf 493}, 223 (1997) 

\bibitem{MN} G. Grinstein, M.A. Mu{\~{n}}oz and Y. Tu, Phys. Rev. Lett.
{\bf 76}, 4376 (1996).

\bibitem{MN2} Y. Tu, G. Grinstein and M.A. Mu{\~{n}}oz,
Phys. Rev. Lett. {\bf 78}, 274 (1997).

\bibitem{mun}W. Genovese, M. A. Mun\~oz,
Phys. Rev. E {\bf 60}, 69 (1999)

\bibitem{first}R. M\"uller, K. Lippert, A. K\"uhnel, and U. Behn, 
Phys. Rev. E {\bf 56}, 2658, (1997)

\bibitem{bar}
A. L. Barab\'{a}si,  H. E. Stanley,
{\it Fractal Concepts in Surface Growth}
Cambridge University Press, Cambridge, 1995.

\bibitem{gard} C.~W.~Gardiner,
{\it Handbook of Stochastic Methods}, Springer Verlag,
Berlin and Heidelberg, 1985

\bibitem{shi}M.~Shiino,
Phys. Rev. A {\bf 36}, 2393, (1987)

\bibitem{heun} P.~E.~Kloeden, E.~Platen, {\it Numerical Solution of 
Stochastic Differential Equations}, Springer Verlag, Berlin, 1992

\bibitem{heun2} M.~San Miguel, R.~Toral, {\it Stochastic Effects in 
Physical Systems}, in {\it Instabilities and Non-Equilibrium Structures}, 
VI, E.~Tirapegui, W.~Zeller Eds., Kluwer Academic Pub., 1997

\end{thebibliography}
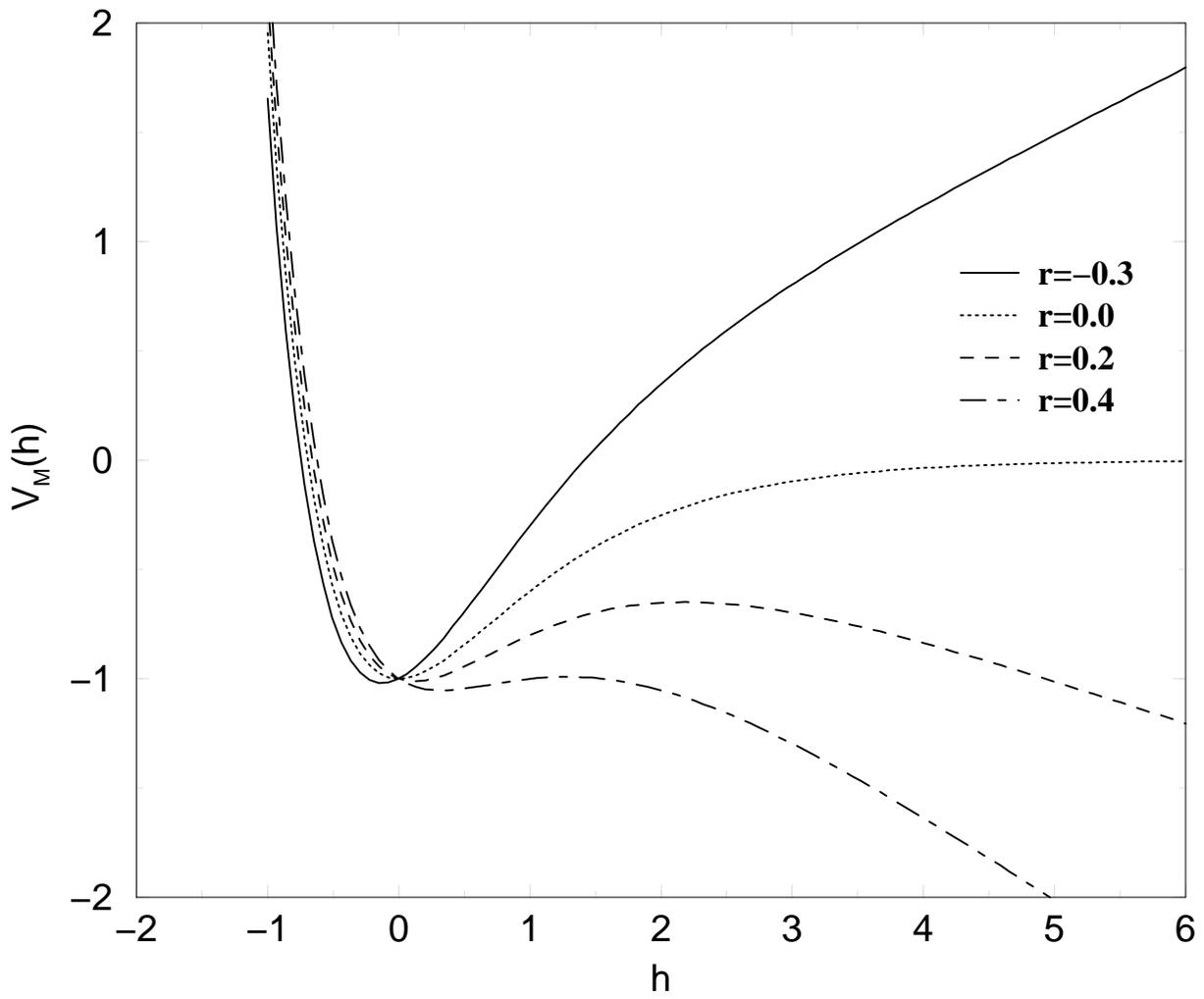
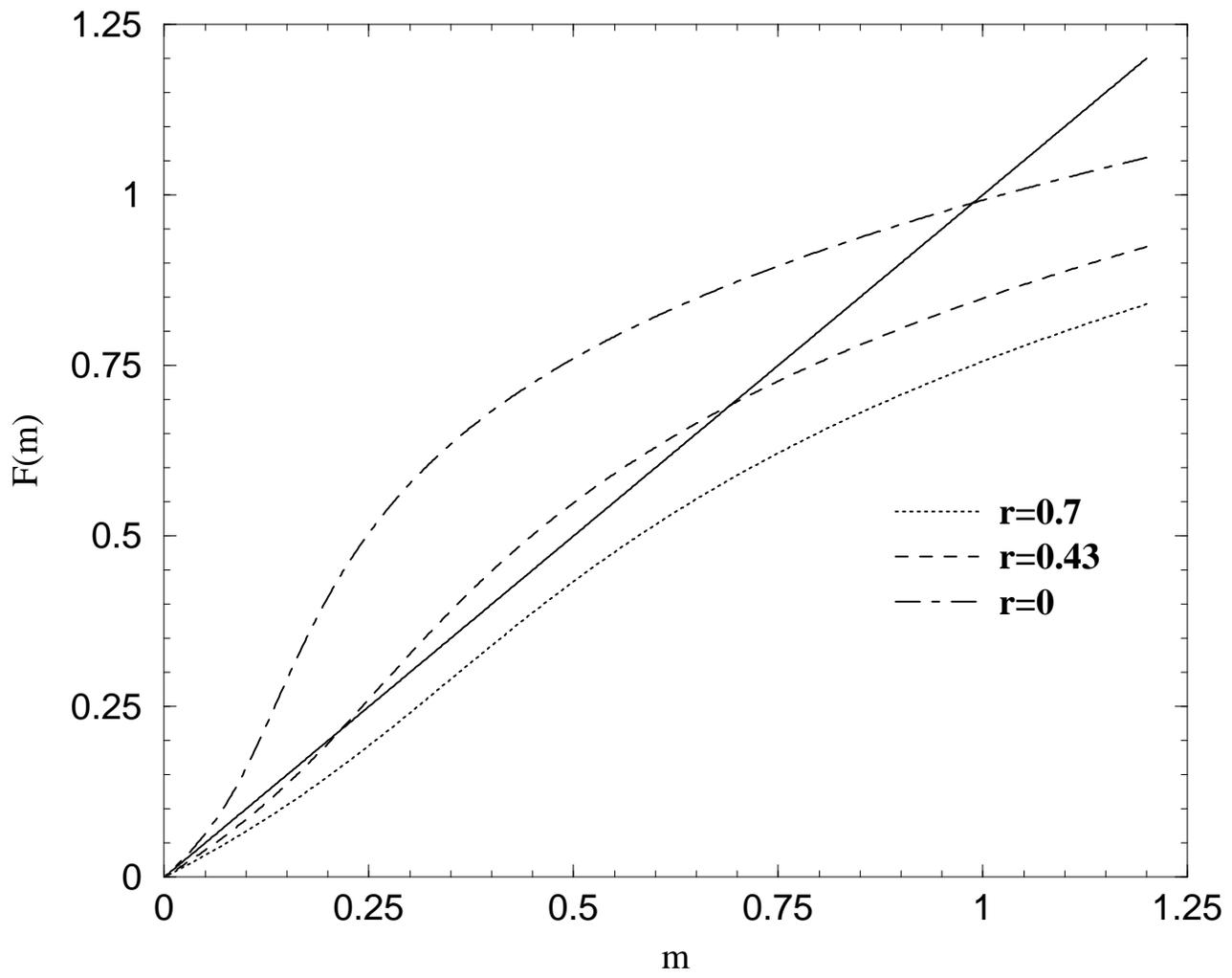
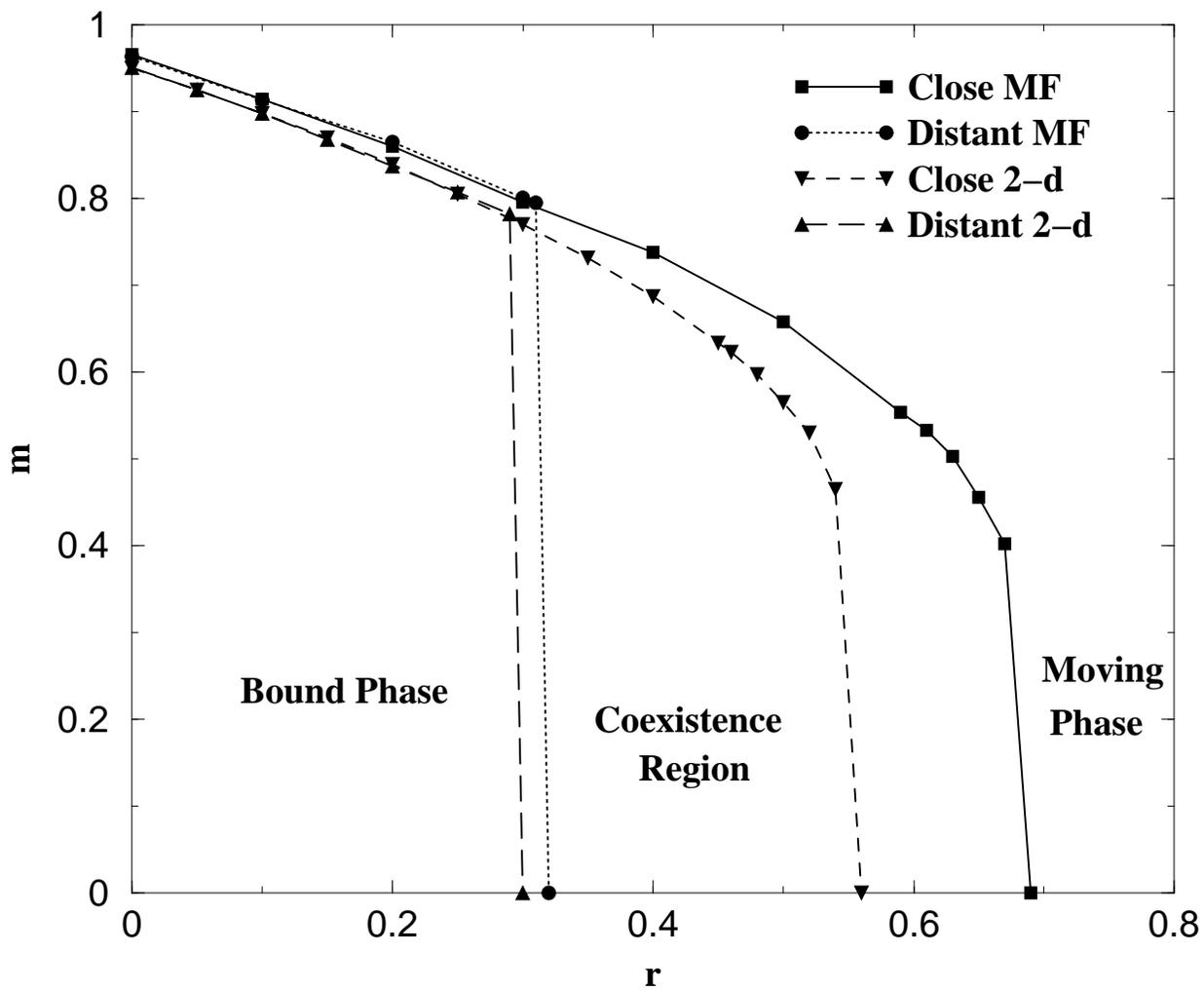
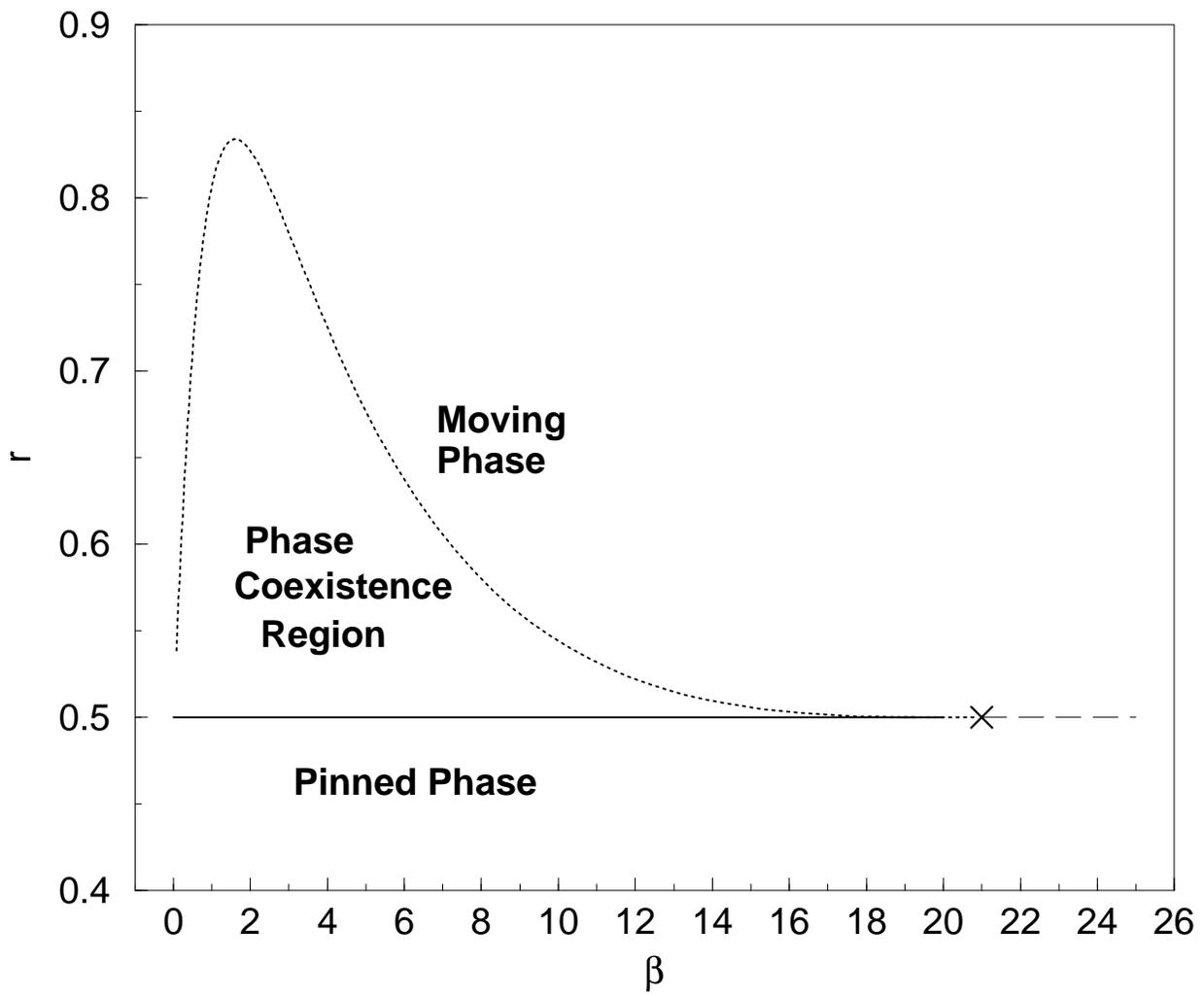
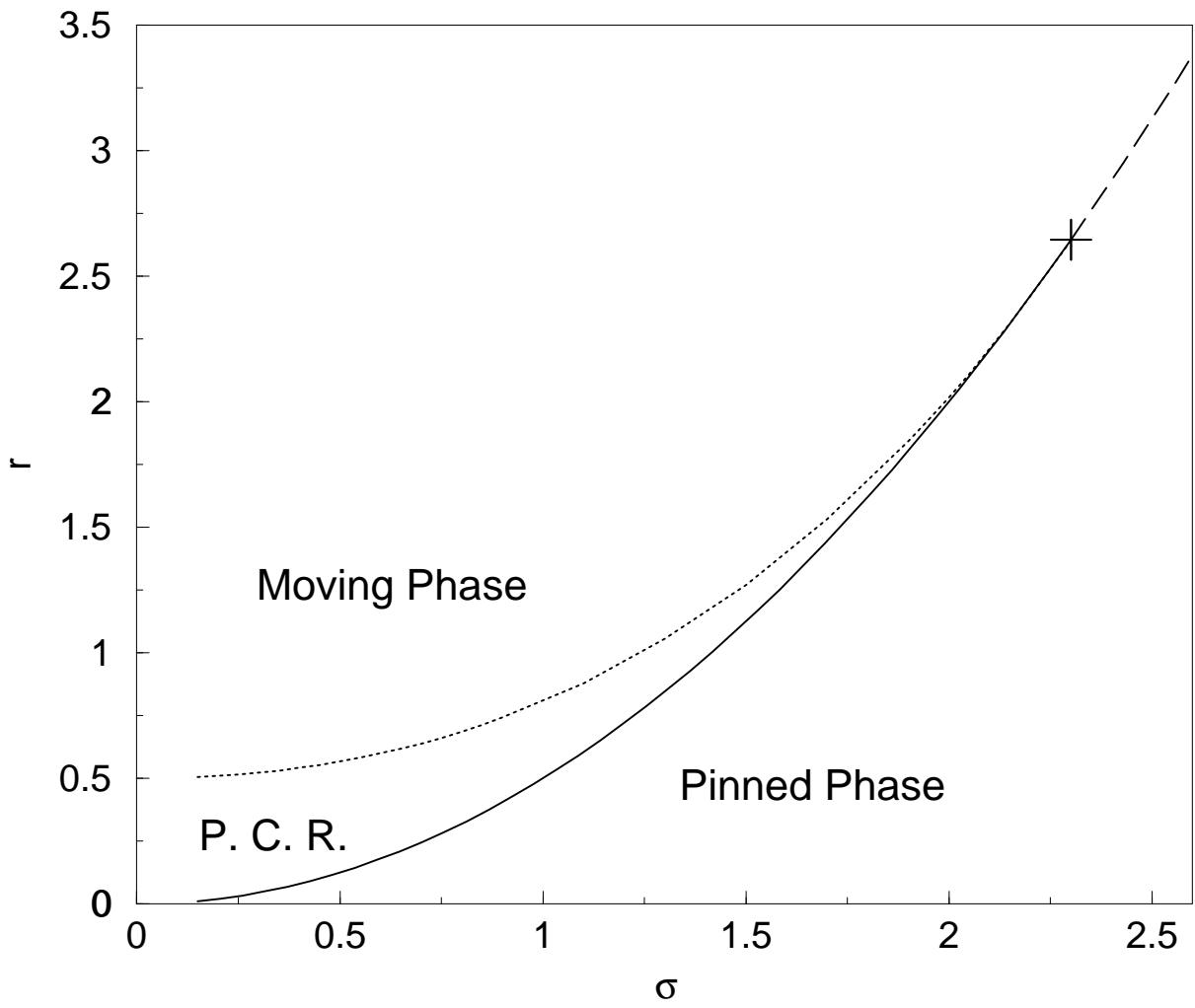
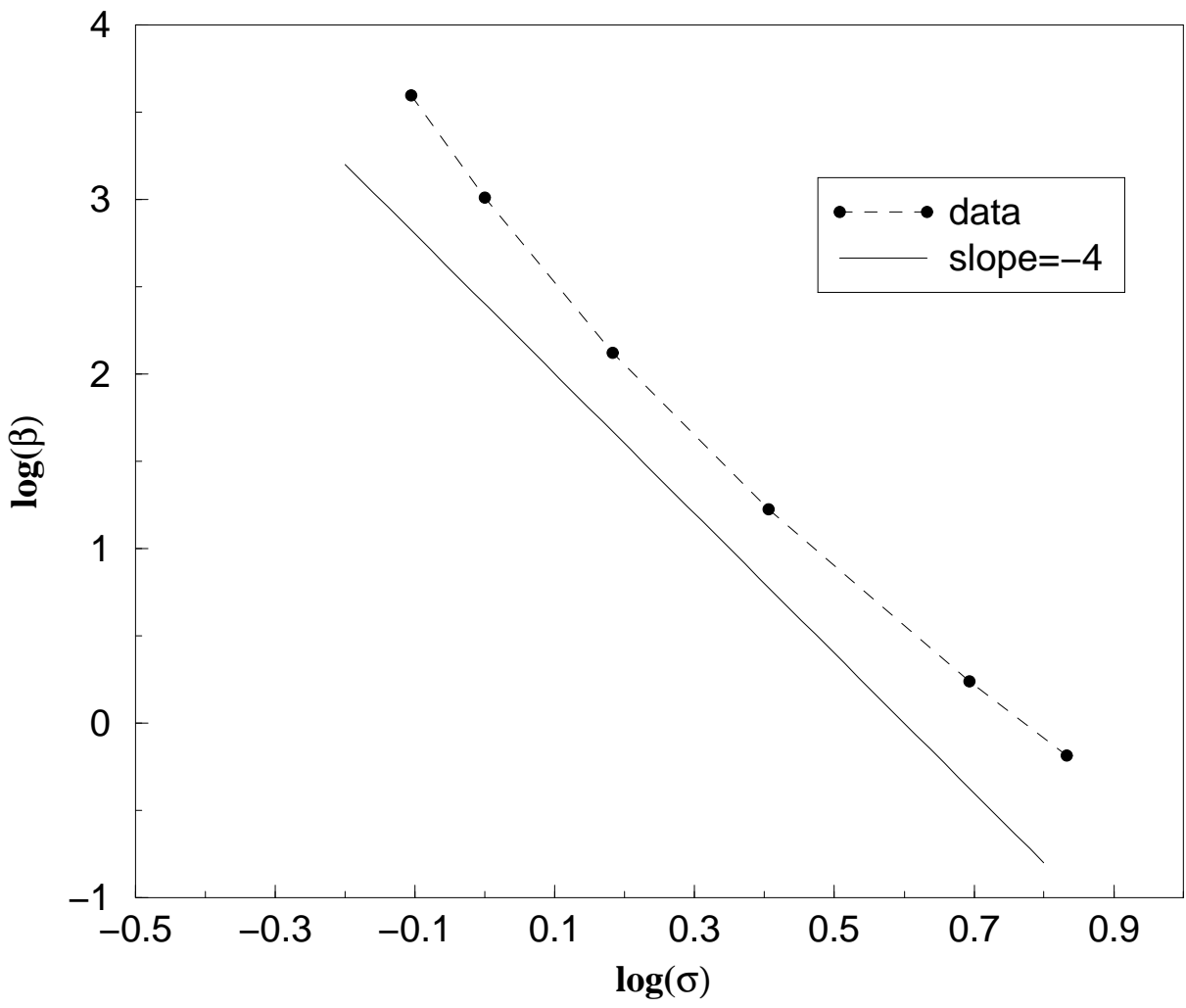
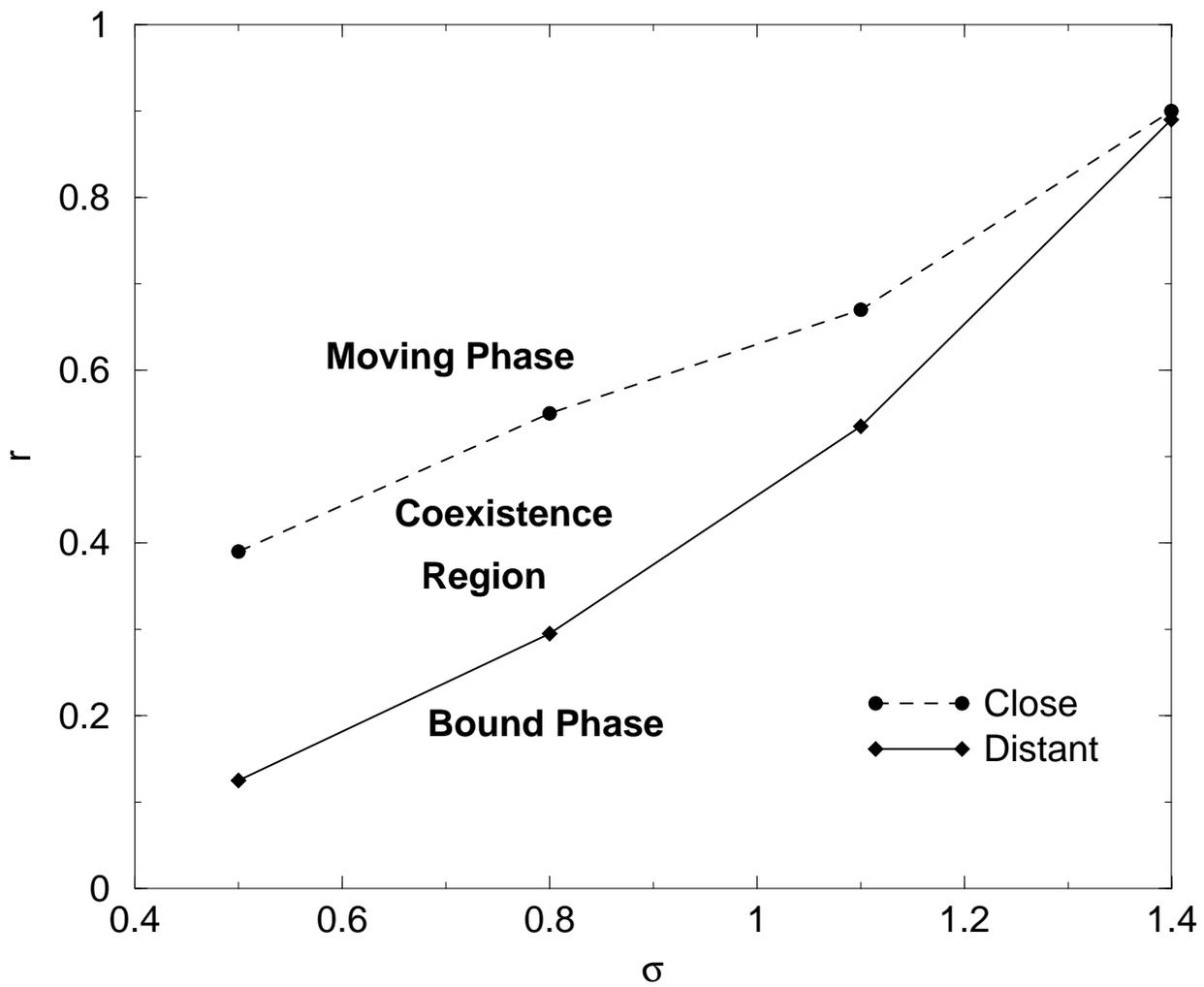
\end{document}